\documentclass[journal=nalefd, manuscript=communication, layout=twocolumn]{achemso}
\usepackage[version=3]{mhchem} 

\usepackage{tabu}
\usepackage{gensymb}
\usepackage{enumitem,kantlipsum}
\usepackage{amsmath}
\usepackage{amssymb}
\usepackage{graphicx}
\usepackage{float}
\usepackage{mhchem}
\usepackage{xcolor}
\usepackage{braket}

\author{Prakriti P. Joshi}
\affiliation[contribution]{These authors contributed equally to this work.}
\alsoaffiliation[University of Chicago, JFI]
{James Franck Institute, University of Chicago, Chicago, IL, 60637 USA}
\author{Ruiyu Li}
\affiliation[contribution]{These authors contributed equally to this work.}
\alsoaffiliation[University of Chicago, Chem]
{Department of Chemistry, University of Chicago, Chicago, IL, 60637 USA}
\alsoaffiliation[University of Chicago, JFI]
{James Franck Institute, University of Chicago, Chicago, IL, 60637 USA}
\author{Joseph L. Spellberg}
\affiliation[University of Chicago, Chem]
{Department of Chemistry, University of Chicago, Chicago, IL, 60637 USA}
\alsoaffiliation[University of Chicago, JFI]
{James Franck Institute, University of Chicago, Chicago, IL, 60637 USA}
\author{Liangbo Liang}
\email{liangl1@ornl.gov}
\affiliation[ORNL]{Center for Nanophase Materials Sciences, Oak Ridge National Laboratory, Oak Ridge, TN, 37830 USA}
\author{Sarah B. King}
\email{sbking@uchicago.edu}
\affiliation[University of Chicago, Chem]
{Department of Chemistry, University of Chicago, Chicago, IL, 60637 USA}
\alsoaffiliation[University of Chicago, JFI]
{James Franck Institute, University of Chicago, Chicago, IL, 60637 USA}

\title{Nano-imaging of the edge-dependent optical polarization anisotropy of black phosphorus}

\abbreviations{}

\begin{document}

\begin{tocentry}
\includegraphics{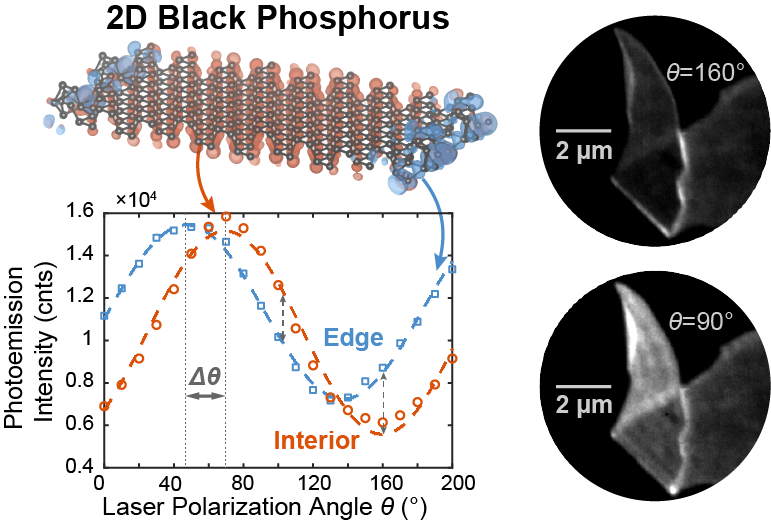}
\end{tocentry}

\begin{abstract}
The electronic structure and functionality of 2D materials is highly sensitive to structural morphology, opening the possibility for manipulating material properties, but also making predictable and reproducible functionality challenging. Black phosphorus (BP), a corrugated orthorhombic 2D material, has  in-plane optical absorption anisotropy critical for applications such as directional photonics, plasmonics, and waveguides. Here, we use polarization-dependent photoemission electron microscopy to visualize the anisotropic optical absorption of BP with 54 nm spatial resolution. We find the edges of BP flakes have a shift in their optical polarization anisotropy from the flake interior due to the 1D confinement and symmetry reduction at flake edges that alter the electronic charge distributions and transition dipole moments of edge electronic states, confirmed with first-principles calculations. These results uncover previously hidden modification of the polarization-dependent absorbance at the edges of BP, highlighting the opportunity for selective excitation of edge states of 2D materials with polarized light.
\end{abstract}

\section{Introduction}
\begin{figure*}[ht]
\centering
		\includegraphics[width=6.5in]{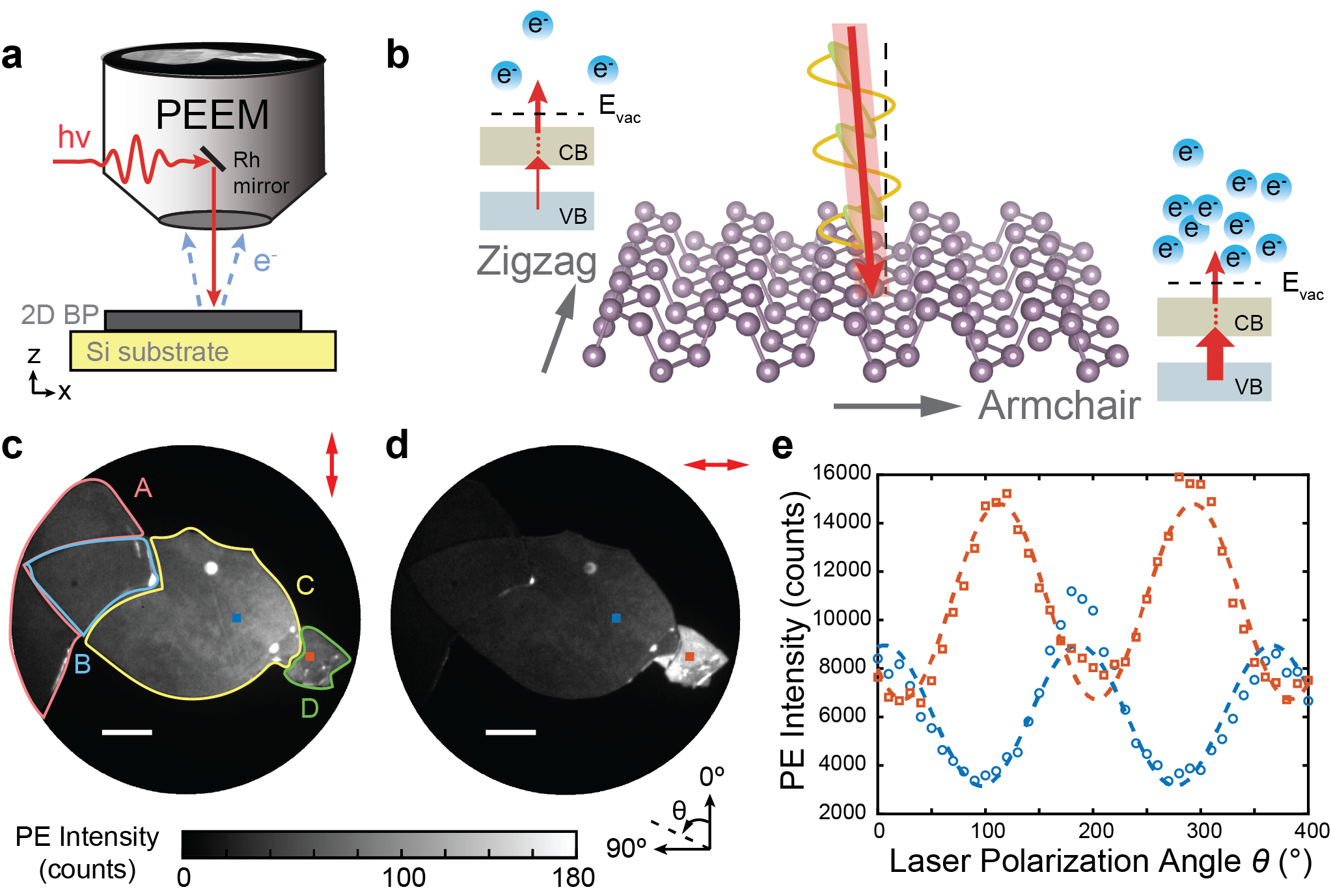}
	\caption{(a) Schematic of NNI PEEM. (b) Crystal structure of monolayer BP crystal. Optical absorption is higher when laser polarization is parallel with armchair direction, leading to increased photoemission efficiency, as shown in the energy diagrams of nPPE processes. (c) PEEM images illuminated by 2.4 eV laser at 0$^{\circ}$ and (d) 90$^{\circ}$ indicated by the red arrows in the upper right corners; Scale bar: 5 $\mu$m. The laser polarization $\theta$ is rotated counter clockwise and shown in the inset. (e) Intensity of the red and blue integrated regions shown by the squares in (c) and (d) as a function of $\theta$ fit to equation \ref{eq:cossquare} (dashed lines).}
	\label{fig:ExperimentDes}
\end{figure*}

Black phosphorus (BP) is an allotrope of elemental phosphorus with a layered crystal structure, high carrier mobility rivaling that of graphene, and a thickness-dependent band gap spanning the visible to the mid-infrared.\cite{Carvalho.Neto.2016} With a corrugated orthorhombic crystal structure, BP also has strong in-plane structural, electronic, and optical anisotropy along the two principal in-plane crystal directions, armchair (AC) and zig-zag (ZZ), shown schematically in Fig. \ref{fig:ExperimentDes}(b).\cite{Ling.Dresselhaus.2015} Similar to other emerging corrugated orthorhombic materials, particularly GeSe, the symmetries and dispersion of the BP conduction and valence bands cause the dielectric function and conductivity tensor for BP to vary significantly along these two crystal directions.\cite{Li.Appelbaum.2014, Low.Neto.2014, Low.Guinea.2014} As a result, optical absorption is highly dependent upon the direction of the incident electric field, and plasmons and polaritons of BP are predicted to be highly directional and possibly hyperbolic.\cite{Li.Appelbaum.2014, Low.Neto.2014, Veen.Yuan.2019} These anisotropic optical, plasmonic, and polaritonic properties of BP make it promising for the development of directional waveguides, plasmonic devices, and light emitters, as well as polarization-dependent photodetectors and thermoelectrics.\cite{Bullock.Javey.2018, Yuan.Cui.2015, Fei.Yang.2014is3, Huber.Huber.2017, Low.Koppens.2016, Biswas.Atwater.2021} 

Morphological features have been observed or predicted to modify the electronic and phononic properties of BP on the nanoscale.\cite{San-Jose.Prada.2016, Liang.Pan.2014, Quereda.Castellanos-Gomez.2016, Ribeiro.Matos.2016} Like other 2D materials prepared via mechanical exfoliation from polycrystalline bulk, few-layer BP can have uncontrolled structural morphology such as grain boundaries, variations in layer thickness, edges, defects, and strain that varies over 10s to 100s of nanometers, which interrupt the properties predicted for defect-free lattices of BP.\cite{San-Jose.Prada.2016, Wei.Long.2018, Liang.Pan.2014, Quereda.Castellanos-Gomez.2016, Ribeiro.Matos.2016, Surrente.Plochocka.2016p8, Shi.Dou.2019, Raja.Chernikov.2019, Zhu.Huang.2018, Kang.Hersam.2015} While some structural morphology can be mitigated by ``bottom-up" preparation methods,\cite{Wu.Hao.2021} morphological features such as edges are omnipresent in many functional applications. In BP nanoribbons, predicted to support anisotropic plasmons and surface plasmon polaritons, edge effects could readily dominate the system's behavior.\cite{Watts.Howard.2019, Das.Drndic.2016, Liu.Aydin.2016, Han.Chen.2018} Edge reconstructions of BP occur readily due to the corrugated orthorhombic lattice and are associated with unique in-gap edge electronic states and phonon modes;\cite{Yao.Jin.2021, Liang.Pan.2014} metallic edge states of BP have been predicted theoretically.\cite{Peng.Wei.2014, Guo.Zeng.2014} However, distinguishing the critical interplay of edge and interior electronic behaviors and their effect on the direction-dependent dielectric function and optical properties of BP is not accessible with the limited spatial resolution of near-IR and visible optical microscopies, and is currently unknown.

Here we use  photoemission electron microscopy (PEEM) to probe the morphology-dependent polarized light absorption of BP with 54 nm spatial resolution, a 4-8x improvement over the spatial resolution of near-IR and visible optical microscopies. PEEM circumvents the optical diffraction limit by imaging the electrons emitted from a material by light.\cite{Dabrowski.Petek.2020} By using two or more photons for photoemission, instead of one, the contrast observed with PEEM images reflects not only the occupied electronic structure and material work function but also the normally unoccupied electronic structure and optical selection rules for optical absorption, similar to how two-photon photoemission spectroscopy probes the unoccupied electronic structure and dynamics of materials.\cite{Ueba.Gumhalter.2007} In contrast to other electron microscopy techniques such as SEM and TEM, no harsh electron beam is required for PEEM, and all of the advantageous properties of light for probing a material (well-resolved photon energies, few femtosecond pulse duration, facile manipulation of focusing and polarization with light optics) are maintained. PEEM with tabletop laser sources has been used previously to image the dynamics of plasmonic fields at metal/vacuum interfaces,\cite{Dabrowski.Petek.2020, Crampton.El-Khoury.2019} ultrafast dynamics in halide perovskites and at p/n junctions,\cite{Doherty.Stranks.2020, Man.Dani.2016} and the packing and alignment of polymers,\cite{Neff.Siefermann.2017} to name a few.

In this paper we show that the edges of black phosphorous flakes have a pronounced difference in their polarization-dependent absorption compared to the main body of a BP flake, displaying $\pm\ 20^\circ$ shift in the polarization angle associated with maximum absorption and photoemission intensity. Through first-principles density functional theory (DFT) calculations, we attribute the edge shift to modification of the electronic charge distributions, and subsequently the optical selection rules, in the near-edge region. Edge-specific optical absorption anisotropy could provide a way to selectively excite the edges of BP, tuning the distribution of charge carriers on the nanoscale even with unfocused light. The reduction in electronic state symmetry that causes the edge-specific absorption in black phosphorus suggests that edge states and properties could be exploited in a wider range of 2D materials, particularly in the design of devices using emerging corrugated orthorhombic 2D materials such as GeSe, arsenene, and GeS.

\section{Methods}
\subsection{Sample Preparation} Few-layer BP is mechanically exfoliated onto a Si substrate with an approximately 2 nm thick native oxide layer\cite{Bohling.Sigmund.2016} in a glove box under \ce{N2} atmosphere. The BP samples are transferred into ultra-high vacuum ($<$ 20 s exposure to ambient conditions) and investigated with PEEM under ultra-high vacuum conditions (UHV, $10^{-10}$ mbar). AFM and Raman microscopy, found in the Supporting Information, confirm the samples are few-layer BP with thicknesses ranging from $\sim$4 nm to 58 nm (8 to 116 layers) and characteristic Raman peaks, A$_{g}^1$, B$_{2g}$, and A$_{g}^2$.\cite{Ling.Dresselhaus.2016} Further details are described in the Supporting Information.
\subsection{Polarization dependent photoemission electron microscopy} Fig. \ref{fig:ExperimentDes}(a) shows a schematic of the experiment. Linearly polarized laser light is directed at near-normal incidence (NNI) via a Rh mirror onto a BP sample in a UHV PEEM microscope chamber (FOCUS GmbH). The angle of incidence is $4^{\circ}$ from normal, allowing the polarization of the laser to be in-plane with respect to the sample at all polarizations. The laser polarization is rotated with a $\lambda/2$ waveplate outside of the UHV chamber, rotating the laser electric field in the sample plane to different angles $\theta$, depicted in Fig. \ref{fig:ExperimentDes}(d). The laser photon energies used in this experiment are 1.55 eV (800 nm) and 2.4 eV (515 nm). These photon energies require 3-photon and 2-photon photoemission (3PPE and 2PPE, Fig. S5) to overcome the $>4.0$ eV work function of few-layer BP.\cite{Cai.Zhang.2014} The first photon excites an electron across the band gap of BP ($\approx 0.3$ eV for $>$ 5 layers) and the subsequent photon(s) within the same laser pulse photoionize the electron, as shown in in Fig. \ref{fig:ExperimentDes}(b). Time-resolved polarization-dependent experiments, described in the Supporting Information, confirm that the polarization-dependent photoemission intensity is due to the across band gap absorption rather than the subsequent photo-ionization photon(s). Photoemitted electrons are accelerated and steered by a set of electron lenses in the PEEM, amplified by a double microchannel plate (MCP)/phosphor screen detector, and imaged by a time-integrated CCD camera. A PEEM image of BP is shown in Fig. \ref{fig:ExperimentDes}(c).
\subsection{Theory}
Plane-wave DFT calculations were carried out using the Vienna \textit{ab initio} simulation package (VASP, version 5.4.4)\cite{PhysRevB.54.11169}. The projector augmented wave (PAW) pseudopotentials were used with a cutoff energy of 400 eV, and the Perdew-Burke-Ernzerhof (PBE) exchange-correlation functional was used.\cite{PhysRevLett.77.3865} To study optical transitions from the BP edges, a monolayer BP nanoribbon with a (1,3) reconstructed edge was selected.\cite{Liang.Pan.2014} A vacuum region of 14 $\text{\AA}$ in the $y$ and $z$ directions were used to avoid spurious interactions with replicas. The whole structure was optimized until the residual forces were below 0.02 eV/$\text{\AA}$ with a $\Gamma$-centered k-point sampling of $20\times1\times1$. After the structural relaxation, we computed the electronic band structure, and the electronic wavefunctions corresponding to each band at each k-point to be post-processed by the VASPKIT code for obtaining the transition dipole moments.\cite{VASPKIT}
\section{Results and Discussion}
\begin{figure*}[ht]
    \centering
		\includegraphics[width=6.5 in]{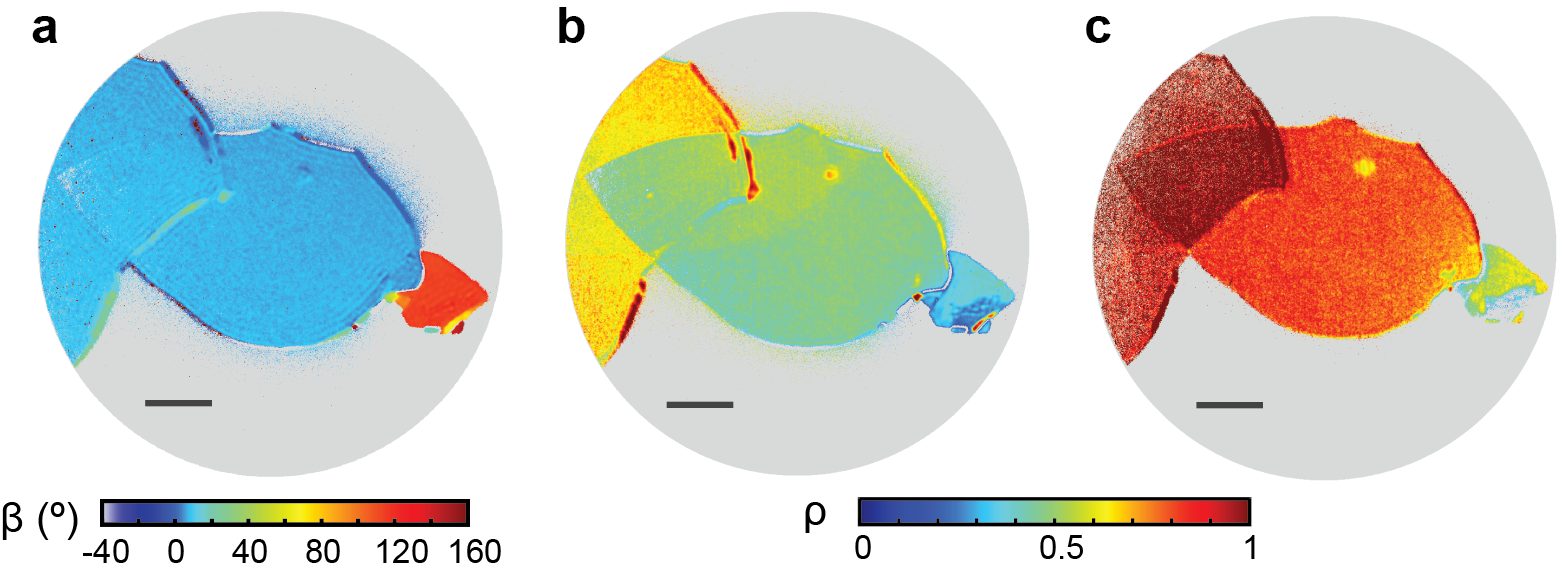}
	\caption{(a) 2.4 eV $\beta$ mapping. (b) 2.4 eV $\rho$ mapping. (c) 1.55 eV $\rho$ mapping. All maps are median-filtered with a 3x3 pixel neighborhood and only pixels with an $R^2>0.6$ are shown. Scale bar: 5 $\mu$m.}
	\label{fig:betarhomapping}
\end{figure*}
Figs. \ref{fig:ExperimentDes}(c) and \ref{fig:ExperimentDes}(d) show PEEM images of the BP flake taken with 2.4 eV illumination and the angle of polarization, $\theta$ (shown schematically in Fig. \ref{fig:ExperimentDes}(d)), set to $0^{\circ}$ and $90^{\circ}$, respectively. The intensity of the PEEM images are all corrected for the polarization-dependent reflectivity of the Rh NNI mirror, as described in the Supporting Information and by Neff \textit{et. al.}\cite{Neff.Siefermann.2017} Four general regions are identified in Fig. \ref{fig:ExperimentDes}(c). Region A is $\approx$ 80 layers thick, B is $\approx$ 116 layers at the overlap between A and C, C is $\approx$ 36 layers, and D is $\approx$ 8 layers, as determined by AFM.

By taking a series of PEEM images as a function of $\theta$, we map out the nanoscale polarization dependence of BP; movies of the polarization-dependent PEEM images can be found in the Supporting Information. Fig. \ref{fig:ExperimentDes}(e) shows the polarization dependence for the integrated photoemission intensity of the 10-by-10 pixel blue and red squares marked on Fig. \ref{fig:ExperimentDes}(c) and (d) as a function of $\theta$ in increments of $10^{\circ}$. The photoemission responses of both regions are periodic with respect to $\theta$ and phase-shifted with respect to each other. The polarization-dependent photoemission response is fit well by a cosine-squared fit of the form
\begin{equation}\label{eq:cossquare}
I_{BP}(\theta)=A\cos^2(\theta-\beta)+C
\end{equation}
where A is the amplitude of the modulation, $\beta$ is the phase shift, and C is the baseline offset of the fit.\cite{Neff.Siefermann.2017} These fits are shown in Fig. \ref{fig:ExperimentDes}(e) by the red and blue dashed lines. The goodness of fit is evaluated by the $R^2$, which in Figure \ref{fig:ExperimentDes}(e) is 0.86 and 0.95 for the blue and red regions, respectively. We evaluate the magnitude of the polarization anisotropy by the dichroism,
\begin{equation}\label{eq:dichroism}
    \rho = \frac{I_{max}-I_{min}}{I_{max}+I_{min}} = \frac{A}{A+2C}.
\end{equation}

\begin{figure}[ht]
    \centering
		\includegraphics[width=3.3 in]{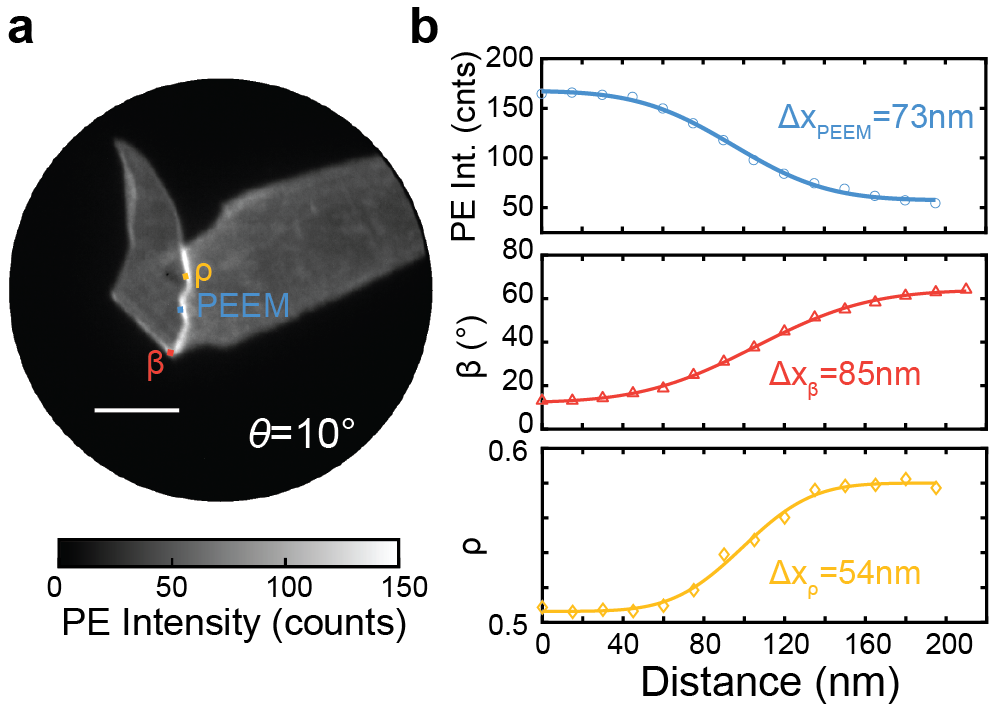}
	\caption{(a) 2.4 eV laser-illuminated PEEM image, taken at $\theta=10 ^\circ$, where the yellow, blue and red line cuts were taken from the PEEM image, $\beta$ map and $\rho$ map, respectively. Scale bar: 3 $\mu$m. (b) Scatter plots show the line profiles from each image or map, fitted with an error function.}
	\label{fig:resolution}
\end{figure}

\textbf{Pixel-by-Pixel Mapping} In order to evaluate the spatial variation of the polarization anisotropy, we perform pixel-by-pixel analysis on the polarization-dependent PEEM images. $I(\theta)$ for each pixel is normalized, fit with equation \ref{eq:cossquare}, and the resulting $\beta$ and $\rho$ values are extracted and calculated for each pixel. Fig. \ref{fig:betarhomapping}(a) shows the resulting $\beta$ map for 2.4 eV excitation. Maps of $\rho$ for 2.4 eV and 1.55 eV are shown in Figs. \ref{fig:betarhomapping}(b) and \ref{fig:betarhomapping}(c), respectively. All maps are median-filtered with a 3-by-3 pixel neighborhood to improve signal-to-noise, and only pixels with a goodness-of-fit $R^2 > 0.6$ are shown. The goodness-of-fit and unfiltered maps can be found in the Supporting Information for comparison. The spatial resolution of Fig. \ref{fig:betarhomapping}(b) is approximately 120 nm, limited by the comparatively large 35 $\mu$m field of view. Fig. \ref{fig:resolution} shows line cuts of a PEEM image, $\beta$ map, and $\rho$ map (Fig. S9) from a flake with a smaller field of view. The reported resolution is the width between 16$\%$ and 84$\%$ of the error function. The smallest features we can measure in this image are 73 nm, 85 nm, and 54 nm for the raw PEEM images, the $\beta$ maps, and the $\rho$ maps respectively and can likely be improved as the spatial resolution of the instrument is approximately 30 nm. This is an approximately 4-8x improvement over conventional visible and near-IR optical microscopy.\cite{Hohenester.Hohenester.2020} 

\textbf{Interpretation of $\beta$ and $\rho$} The $\beta$ values for the interiors of regions A, B, and C are qualitatively similar, however, region D is phase shifted by $-50^{\circ}$ to $80^{\circ}$ compared to A, B, and C (Fig. \ref{fig:betarhomapping}(a)). The anisotropic optical response of BP is well known; for $h\nu$ $<$ 3 eV the absorption coefficient for an electric field polarized along the armchair (AC) direction of the lattice exceeds that of the zigzag (ZZ) direction by $\sim1-2$ orders of magnitude.\cite{Ling.Dresselhaus.2016} Therefore, more electrons will be excited across the band gap when the laser polarization is aligned with the armchair axis of the BP flake, resulting in the highest photoemission intensity, shown schematically in Fig. \ref{fig:ExperimentDes}(b). From equation \ref{eq:cossquare}, therefore, $\theta = \beta$ is the angle where the laser polarization is aligned with the armchair axis. As the $\beta$ values of A, B, and C are approximately the same, we conclude that these are regions of different thicknesses of the same crystalline piece rather than regions with mismatched lattice parameters. However, the measured $\beta$ of the majority of region D is rotated with respect to A, B, and C by approximately 50 degrees, suggesting it is non-contiguous with the rest of the flake. Comparison to the corresponding AFM image (Figs. S1(a)-(c)) indeed shows that region D is a flake broken off from the A, B, and C regions with folded and/or overlapping areas at the bottom of region D. While the $\beta$ maps are largely photon energy independent, comparison of Figs. \ref{fig:betarhomapping}(b) and \ref{fig:betarhomapping}(c) shows that the dichroism maps, $\rho$, are photon energy dependent. The dichroism, $\rho$, is a measure of the contrast between the photoemission intensity at AC (maximum intensity) and at ZZ (minimum intensity), reflecting the differential optical absorbance between the AC and ZZ directions for a particular BP thickness and excitation energy. This differential optical absorption varies as a function of layer thickness and photon energy,\cite{Ling.Dresselhaus.2016, Lan.Cai.2016} explaining the difference in the $\rho$ maps for 1.55 and 2.4 eV.
\begin{figure}[ht]  
        \centering 
		\includegraphics[width=3in]{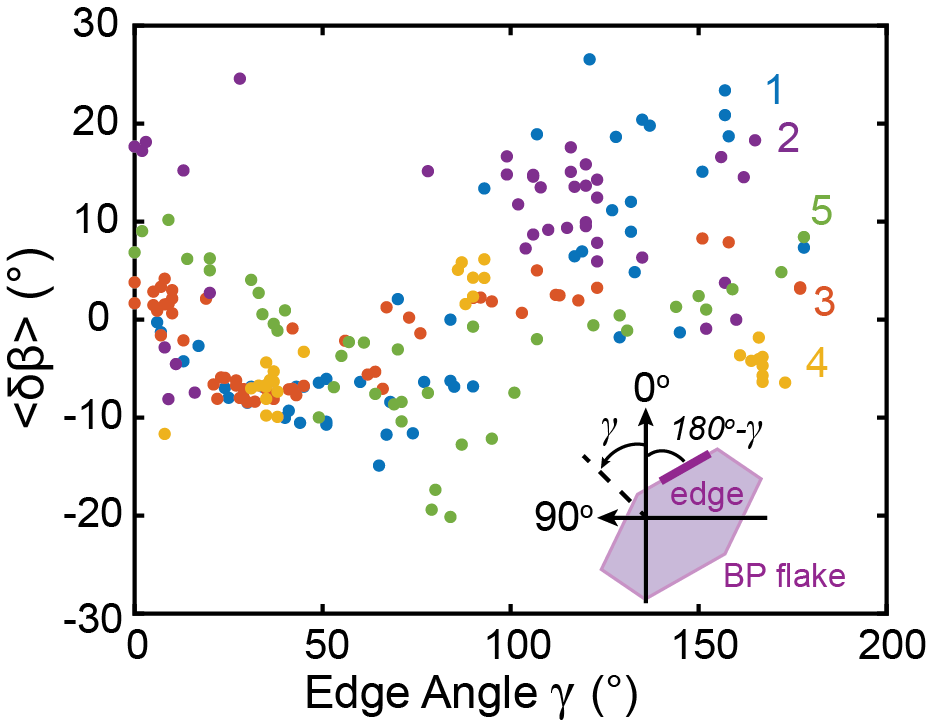}
	\caption{Average phase shift $\langle\delta\beta\rangle$ as a function of edge orientation $\gamma$, shown schematically in the inset, for five different black phosphorus flakes.}
	\label{fig:OrientationalDependence}
\end{figure}

\textbf{Edge-dependent optical anisotropy} At both photon energies (Fig. \ref{fig:betarhomapping}(a) and Fig. S7), the edges of regions A, B, and C have $\beta$ values that are phase shifted by approximately $-20^{\circ}$ to $20^{\circ}$ relative to the interiors of the flakes, where the armchair direction is $\beta=7^\circ$. Edge-dependent phase shifts are not readily observed in region D, which is significantly thinner (8 layers) than regions A-C, nor are they observed at monolayer edges intentionally introduced through sublimation (Fig. S18).\cite{Fortin-Deschenes.Moutanabbir.2016, Liu.Hersam.2015, Kumar.Heun.2018} Further information regarding the intentionally created edges can be found in the Supporting Information.

Not all edges of the flake have the same phase shift compared to the armchair direction, but along an entire edge segment the phase shift remains predominantly the same. These phase shifts are reproducible across different BP flakes and samples. We analyzed the average phase shift of the edge from the flake interior ($\langle\delta\beta\rangle$) for short line segments versus the angle $\gamma$ of the edge for five different BP flakes ($\beta$ maps showing these segments can be found in Fig. S12) and find a persistent phase shift of up to $\pm\ 20^{\circ}$ at flake edges despite the varied $\gamma$ for different segments and armchair directions of the flake interiors ranging from $7^\circ$ to $156^\circ$ (Fig. \ref{fig:OrientationalDependence}).

The phase-shifted optical anisotropy of the edges of regions  A, B, and C is independent of folds, wrinkles, and oxides as verified by AFM (Fig. S1) and experiments performed before and after annealing of the samples to 350 $^{\circ}$C (Fig. S11), the temperature required to achieve a pristine oxide-free surface.\cite{Liu.Hersam.2015, Kumar.Heun.2018, Edmonds.Fuhrer.2015, Kuntz.Warren.2017} We also exclude near-field modification of the incident EM fields as the source of the edge phase shifts. With near-field modifications we would expect a correlation between the $\beta$ values at flake edges and the angle of the edge defined in the laboratory frame, regardless of the orientation of the AC and ZZ directions of the flake, causing phase shifts to be very large with some flakes and small with others depending on the flake orientation. This is not observed in Fig. \ref{fig:OrientationalDependence}, the phase shifts relative to AC orientation are up to $\pm\ 20^{\circ}$ regardless of the interior flake orientation with respect to the laboratory frame.

\begin{figure*}[h]  
    \centering 
	\includegraphics[width=6.5in]{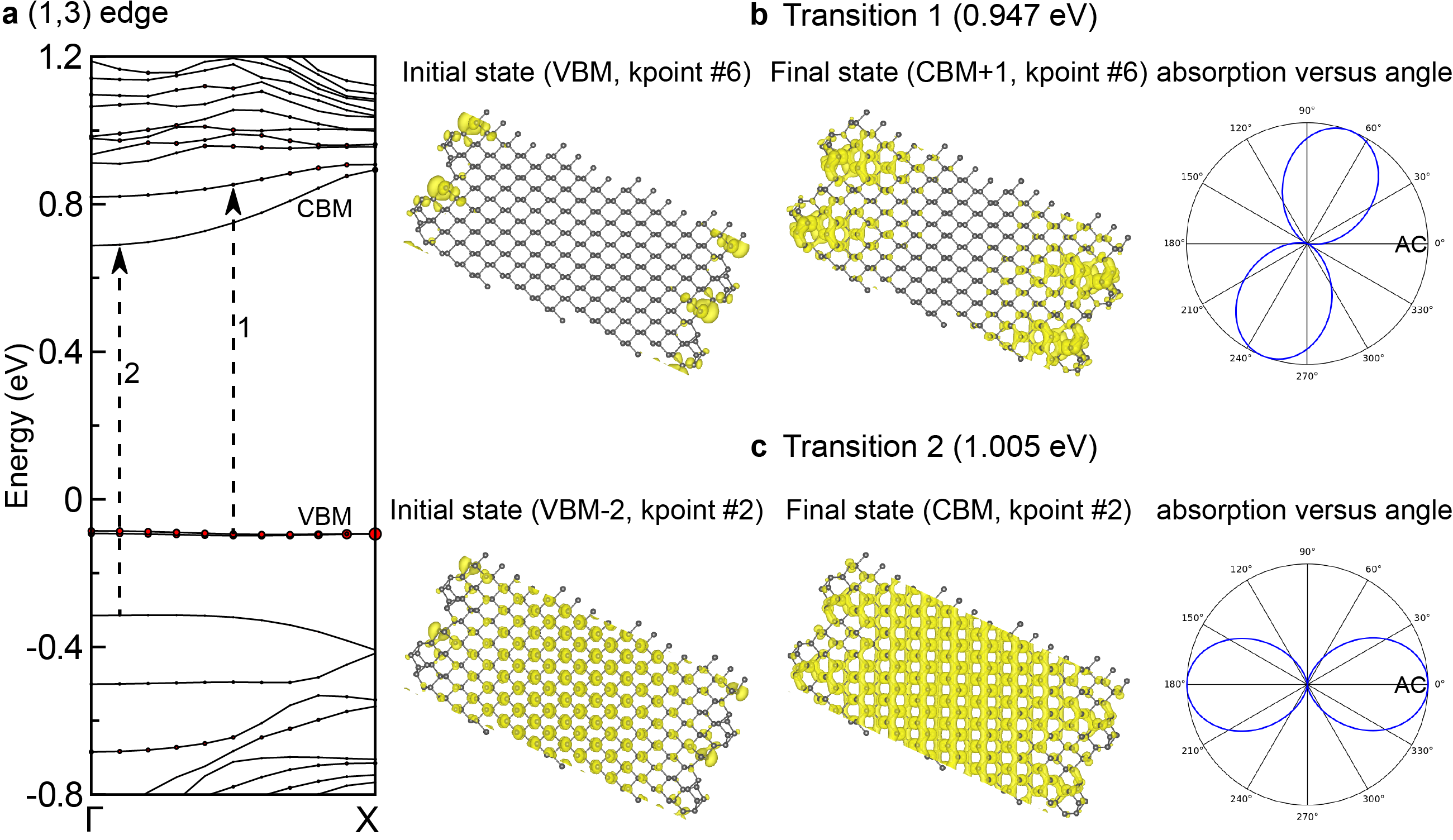}
	\caption{(a) Electronic band structure of a monolayer BP nanoribbon with a (1,3) edge with the valence band maximum (VBM) and conduction band minimum (CBM) labeled where the Fermi level is Energy = 0 eV. Red circles superimposed on the bands correspond to the contribution from the edge atoms. For two representative optical transitions (indicated by the arrows), spatial distributions of the charge densities of the initial and final states, and polarization-angle-dependent absorption profiles are shown in (b) and (c).}
	\label{fig:TheoryFig}
\end{figure*}

To understand the experimentally observed phase-shifted behavior at the edges of BP flakes, we carried out proof-of-principle DFT calculations on a (1,3) reconstructed edge of monolayer BP; calculations on multilayer BP are computationally too expensive. The AC direction is defined as (1,0) while the ZZ direction corresponds to (0,1). Fig.~\ref{fig:TheoryFig}(a) shows the calculated electronic band structure of the BP nanoribbon with the (1,3) edge. By computing the contribution of the edge atoms to each band state (indicated by the size of the red circles), we can identify the two bands at approximately $-0.1$ eV as edge bands.
Two representative transitions with different transition energies, approximately 0.95 eV and 1.0 eV, are indicated by the arrows in Fig. \ref{fig:TheoryFig}(a). It is important to note that because of the limitations of obtaining accurate energies in DFT, the energies and momenta of the transitions are illustrative rather than for quantitative comparison to experiment. Fig. \ref{fig:TheoryFig}(b) and (c) show the spatial distributions of charge densities of initial valence state and final conduction state for the two transitions. In transition 1, shown in Fig. \ref{fig:TheoryFig}(b), the charge densities of the initial (VBM) and final (CBM+1) states are largely confined at the edges of BP that have a symmetry reduction compared to the interior atomic structure. Therefore, their spatial distributions and symmetries are significantly modified compared to those of transition 2, where the charge densities of the initial and final states are largely in the interior of BP, shown in (c). Consequently, the transition dipole moment of transition 1 is notably rotated away from the AC direction, changing the optical selection rules. This is evident in the polarization-angle-dependent optical absorption profile shown on the right of Fig. \ref{fig:TheoryFig}(b) where the angle of maximum absorption is 69$\degree$ (more details regarding the selection rules can be found in the Supporting Information). These results are in stark contrast to typical optical transitions that occur in the interior of BP shown in Fig. \ref{fig:TheoryFig}(c) where the transition dipole moment and the maximum absorption are along the AC direction (i.e., 0.0$\degree$), as expected for 2D BP.\cite{Tran.Yang.2014,Ling.Dresselhaus.2016} In short, our calculations demonstrate that the optical selection rule is modified at the edges of BP due to the 1D confinement and symmetry reduction in comparison with 2D BP, leading to orientation changes of the transition dipole moments at the edges and the experimentally observed phase variations in the maximum absorption direction between the edges and interior. More details for all of the transitions for a range of photon energies can be found in the Supporting Information.

Our calculations are focused on monolayer BP due to the computational cost, and indicate that edge phase shifts should occur in BP monolayers. However, edge phase shifts are not detected in our experiments on 8-layer, or thinner, regions. As discussed in the Supporting Information, transitions originating from the interior often show stronger intensities than edge-related transitions. Even with the excellent spatial resolution of these experiments, there is inevitably a mixture of edge and interior contributions observed at the edges of our images. With extremely thin BP flakes, when accounting for our spatial resolution, the comparative contribution from the edges and the interior leads to very small computed phase shifts in the monolayer BP nanoribbon, on the order of 2 $\degree$. We expect that with increasing flake thickness the contribution from the edges compared to the interior increases, leading to the larger phase shifts that we can detect experimentally. Similar thickness dependence is observed in edge-dependent Raman modes, which appear in thick BP samples but disappear with decreasing thickness due to the signal reduction of edge Raman modes with decreasing thickness.\cite{Ribeiro.Matos.2016}

\section{Conclusion}
In conclusion, we have used photoemission electron microscopy (PEEM) to image the nanoscale variation in the polarization-dependent photoemission response of black phosphorus. Enabled by our 54 nm spatial resolution, we observe that the edges of BP flakes have a phase shift in their polarization-dependent absorption of $\pm\ 20 ^\circ$ compared to the interior of the flake. Through comparison with DFT calculations, we assign these phase shifts to modification of the symmetry of the occupied and unoccupied wavefunctions at and in the vicinity of BP edges, due to the 1D confinement and symmetry reduction of BP edges. The unique absorption properties of BP edges mean that the extinction coefficients and complex dielectric function of BP edges are also unique from flake interiors, determining the functionality of BP in photonics-on-chip, waveguides, and directional plasmonic applications. Edge-specific optical absorption could also enable selective excitation of nanoscale BP edges even with far-field optical excitation, controlling the spatial distribution of excited charge carriers on the nanoscale. This work highlights how structural morphology can modify 2D material properties as simple as optical absorption, providing challenges and opportunities for material control.
 
\begin{acknowledgement}
This work was funded by the Office of Basic Energy Sciences, U.S. Department of Energy (Grant No. DE-SC0021950). This work was partially supported by the University of Chicago Materials Research Science and Engineering Center, which is funded by the National Science Foundation under award number DMR-2011854 and DMR-1420709. This work made use of the shared facilities at the University of Chicago Materials Research Science and Engineering Center, supported by National Science Foundation under award number DMR-2011854. P.P.J. acknowledges support from a MRSEC-funded Kadanoff-Rice fellowship (DMR-2011854 and DMR-1420709). S.B.K. acknowledges start-up funding support from the University of Chicago and the Neubauer Family Assistant Professors Program. Theoretical calculations were conducted by L.L. at the Center for Nanophase Materials Sciences, which is a Department of Energy Office of Science User Facility. L.L. used resources of the Compute and Data Environment for Science (CADES) at the Oak Ridge National Laboratory, which is supported by the Office of Science of the U.S. Department of Energy under Contract No. DE-AC05-00OR22725. 
\end{acknowledgement}

\begin{suppinfo}
Supporting information can be found online including, details of sample preparation, AFM and Raman microscopy characterization of BP flakes, details regarding the Rh mirror reflectivity correction, evaluation of the 2PPE versus 3PPE photoemission processes, time-resolved polarization-dependent data, goodness of fits for the pixel-by-pixel analysis, $\beta$ and $\rho$ maps before median filtering, the $\beta$ maps for the five flakes of Fig. \ref{fig:OrientationalDependence}, polarization dependence of sublimated black phosphorus samples, and DFT calculations. The authors are committed to making research data displayed in publications digitally accessible to the public at the time of publication. All data generated and used in this publication is organized, curated, and made available for exploration to the public using the software Qresp (Curation and Exploration of Reproducible Scientific Papers: \url{http://qresp.org/}).

\end{suppinfo}
\bibliography{BP_Bib_02222022.bib}

\end{document}